\documentclass[preprint,12pt]{elsarticle}

\usepackage{amssymb}
\usepackage{amsmath}
\usepackage{graphics}

\journal{Physics Letters B}

\begin{document}

\begin{frontmatter}

\title{Note on the Evolution of the Gravitational Potential in Rastall Scalar Field Theories}

\author{J.~C.~Fabris, M.~Hamani Daouda, O.~F.~Piattella}
\ead{fabris@pq.cnpq.br, daoudah8@yahoo.fr, oliver.piattella@ufes.br}
\address{Departamento de F\'isica, Universidade Federal do Esp\'irito Santo, avenida Ferrari 514, 29075-910 Vit\'oria, Esp\'irito Santo, Brazil}

%%%%%%%%%%%%%%%%%%%%%%%%%%%%%%%%%%%%%%%%%%%%%%%%%%%%%%%%%%%%%%%%%%%%%%%%%%%%%%%%%%%%%%%%%%%%%%%%%%%%%%%%%%%%%%%%%%%%%

\begin{abstract}
We investigate the evolution of the gravitational potential in Rastall scalar field theories. In a single component model a consistent perturbation theory, formulated in the Newtonian gauge, is possible only for $\gamma = 1$, which is the General Relativity limit. On the other hand, the addition of another canonical fluid component allows to consider the case $\gamma \neq 1$.
\end{abstract}

%\pacs{95.35.+d, 95.36.+x, 98.80.-k, 98.80.Jk}


\end{frontmatter}

\maketitle

%%%%%%%%%%%%%%%%%%%%%%%%%%%%%%%%%%%%%%%%%%%%%%%%%%%%%%%%%%%%%%%%%%%%%%%%%%%%%%%%%%%%%%%%%%%%%%%%%%%%%%%%%%%%%%%

\section{Introduction}

The nature of dark matter and dark energy is one of the most important issues today in physics. There are strong observational evidences in astrophysics and cosmology for the existence of these two exotic components of the cosmic energy budget, indicating that about $95\%$ of the Universe is composed of dark matter (about $25\%$) and dark energy (about $70\%$), but no direct detection has been reported until now. The usual candidates to dark matter (neutralinos and axions, for example) and dark energy (cosmological constant, quintessence, etc.) lead to very robust scenarios, but at same time they must face theoretical and observational issues. For recent reviews on the subject, see for example \cite{li, caldwell, bertone, pad, martin}.
\par
An interesting proposal concerning the nature of dark matter and dark energy are the unification models. According to the latter, the whole dark sector is a manifestation of a single entity. The paradigm is a perfect fluid model called Chaplygin gas \cite{pasquier}, but recently it has also been shown that viscous models may lead to unification scenarios \cite{viscous}. In spite of their great appealing, however, unification models suffer from severe problems when confronted with observations, since the parameter estimations from different tests lead to contradictory values \cite{contra}. One way to surmount this conflict is to encode the unification model in a scalar-tensor framework, the exotic fluid being described by a self-interacting scalar field. It is not easy to implement this idea, since a canonical self-interacting scalar field has a sound speed equal to the speed of light, and it cannot represent dark matter in the past evolution of the universe, as required by the unification program \cite{scalar}.
\par
A scalar model that is able to represent a realization of dark matter and dark energy can be obtained with non-canonical self-interacting scalar fields. One example is a scalar field obeying the structure of Rastall's theory, for which the usual conservation law for the matter content is modified \cite{Rastall:1973nw}. The price to pay is the loss of a Lagrangian formulation, at least in the context of Riemannian geometry. In Rastall's theory a new dimensionless parameter $\gamma$ is introduced, which measures the departure from the usual equations of General Relativity. When $\gamma = 1$, the General Relativity theory is recovered. As in the case of other unification scenarios, the theory is able to satisfactorily reproduce the kinematic background observational tests (e.g. based on type Ia supernovae surveys), but essentially reduces to General Relativity if a hydrodynamical approach is used for the fluid obeying the new conservation laws and the matter power spectrum data are used \cite{eduardo}. However, the agreement improves if a non-canonical scalar field is employed instead of a fluid. Moreover, if $\gamma = 2$ this non-canonical scalar field may behave as dark matter, and may lead to a unification scenario \cite{Gao:2009me}.
\par
Can Rastall's theory pass another important test, represented by the Integrated Sachs-Wolfe effect? In order to answer this question, the gravitational potential $\Phi$ must be analysed. Using the newtonian gauge, we find an astonishing property of Rastall's theory: in a scenario with just one component, given by the non-canonical Rastall scalar field, only homogeneous solutions for $\Phi$ are admitted, unless $\gamma = 1$. But homogeneous solutions for $\Phi$ are not real perturbations, since they can be re-absorbed in the background metric through a suitable reformulation of the background functions. Thus, the case $\gamma = 1$ seems to be forced. This fundamental drawback can be cured if a two-fluid model is formulated: one scalar field obeying the modified conservation laws, and a fluid obeying the canonical conservation law. The main conclusion is: Rastall non-canonical scalar field requires a matter component (baryons, for example) in order for the theory to make sense at the perturbative level.
\par
In the next section, we introduce the Rastall non-canonical scalar field. In section 3, we discuss the sound speed issue in this theory, and in section 4 we investigate the evolution of the gravitational potential and find the constraint $\gamma = 1$. In section 5, we show how to soothe such constraint, by adding a canonical fluid component, and we study the gravitational potential in a special case. In section 6, we present our conclusions.

%%%%%%%%%%%%%%%%%%%%%%%%%%%%%%%%%%%%%%%%%%%%%%%%%%%%%%%%%%%%%%%%%%%%%%%%%%%%%%%%%%%%%%%%%%%%%%%%%%%%%%%%%%%%%%%

\section{Scalar field in Rastall's theory}

In Rastall's theory \cite{Rastall:1973nw} a scalar field $\phi$ is characterised by the following stress-energy tensor:
\begin{equation}\label{TRast}
 T_{\mu\nu} = \phi_{,\mu}\phi_{,\nu} - \frac{2 - \gamma}{2}g_{\mu\nu}\phi_{,\alpha}\phi^{,\alpha} + g_{\mu\nu}(3 - 2\gamma)V(\phi)\;,
\end{equation}
where $\gamma$ is a parameter. When $\gamma = 1$, we recover the corresponding theory in General Relativity.

We consider a spatially flat Friedmann-Lema\^itre-Robertson-Walker metric
\begin{equation}\label{FLRWmet}
 ds^2 = dt^2 - a^2(t)\delta_{ij}dx^idx^j\;,
\end{equation}
and its perturbation written in the newtonian gauge
\begin{equation}\label{pertFLRWmet}
 ds^2 = [1 + 2\Phi(t,x^i)]dt^2 - a^2(t)[1 - 2\Phi(t,x^i)]\delta_{ij}dx^idx^j\;,
\end{equation}
where $\Phi(t,x^i)$ is the gravitational potential.

Consider a scalar field perturbation of the form $\phi(t,x^i) = \phi_0(t) + \delta\phi(t,x^i)$ and derive from Eq.~\eqref{TRast} together with Eqs.~\eqref{FLRWmet} and \eqref{pertFLRWmet} the background and first-order perturbed mixed-component stress-energy tensors. The former is
\begin{eqnarray}
\label{rhoR} {}^{(0)}T^0{}_0 &=& \rho = \frac{\gamma}{2}\dot{\phi_0}^2 + (3 - 2\gamma)V(\phi_0)\;,\\
{}^{(0)}T^0{}_i &=& 0\;,\\
\label{pR} {}^{(0)}T^i{}_j &=& -p\delta^i{}_j = -\left[\frac{2 - \gamma}{2}\dot{\phi_0}^2 - (3 - 2\gamma)V(\phi_0)\right]\delta^i{}_j\;,
\end{eqnarray}
where the dot denotes derivative with respect to $t$ and $\rho$ and $p$ are the scalar field energy density and pressure, respectively. The perturbative quantities have the form
\begin{eqnarray}
\label{deltarho} \delta T^0{}_0 &=& \delta\rho = \gamma\dot{\phi_0}\dot{\delta\phi} - \gamma\Phi\dot{\phi_0}^2  + (3 - 2\gamma)V_{,\phi}\delta\phi\;,\\
\delta T^0{}_i &=& \dot{\phi_0}\delta\phi_{,i}\;,\\
\label{deltap} \delta T^i{}_j &=& -\delta p\delta^i{}_j = \left[(\gamma - 2)\dot{\phi_0}\dot{\delta\phi} + (2 - \gamma)\Phi\dot{\phi_0}^2 + (3 - 2\gamma)V_{,\phi}\delta\phi\right]\delta^i{}_j\;,
\end{eqnarray}
where $,i$ denotes the spatial derivative with respect to the coordinate $x^i$ and $V_{,\phi} := dV(\phi)/d\phi$. The modified Klein-Gordon equation has the covariant form
\begin{equation}\label{KGeqcov}
 \square\phi + (3 - 2\gamma)V_{,\phi} = (1 - \gamma)\frac{\phi^{,\rho}\phi^{,\sigma}\phi_{;\rho;\sigma}}{\phi_{,\alpha}\phi^{,\alpha}}\;,
\end{equation}
where $\square\phi := \phi_{;\alpha}{}^{;\alpha}\;.$ From Eq.~\eqref{KGeqcov}, it appears clearer that $\gamma = 1$ restores the General Relativity case. Insert again $\phi = \phi_0 + \delta\phi$ into Eq.~\eqref{KGeqcov} and employ metric \eqref{FLRWmet} and \eqref{pertFLRWmet}, in order to find
\begin{equation}\label{KGeqcovbg}
 \gamma\ddot{\phi_0} + 3H\dot{\phi_0} + (3 - 2\gamma)V_{,\phi} = 0\;, 
\end{equation}
where $H := \dot{a}/a$ is the Hubble parameter. Employing the conformal time $d\eta = dt/a(t)$ we write
\begin{equation}\label{KGeqcovbgconf}
 \gamma\phi_0'' + (3 - \gamma)\mathcal{H}\phi_0' + (3 - 2\gamma)a^2V_{,\phi} = 0\;, 
\end{equation}
where the prime denotes derivation with respect to the conformal time and $\mathcal{H} := a'/a$. The perturbed modified Klein-Gordon equation has the following form:
\begin{eqnarray}\label{KGeqcovpertconf}
 \gamma\delta\phi'' &+& (3 - \gamma)\mathcal{H}\delta\phi' - \nabla^2\delta\phi\nonumber\\ 
- (3 + \gamma)\phi_0'\Phi' &+& 2(3 - 2\gamma)a^2V_{,\phi}\Phi + (3 - 2\gamma)a^2V_{,\phi\phi}\delta\phi = 0\;,
\end{eqnarray}
where $V_{,\phi\phi} := d^2V(\phi)/d\phi^2$. 

Differently from the $\gamma = 1$ case, in Rastall's theory we have more degrees of freedom. Indeed, considering $T^{\mu}{}_{\nu;\mu} = 0\;:$
\begin{equation}\label{Tmunumu}
 \phi_{,\nu}\left[\square\phi + \left(3 - 2\gamma\right)V_{,\phi}\right] = \left(1 - \gamma\right)\phi^{,\mu}\phi_{;\mu;\nu}\;,
\end{equation}
one notices that the contraction with $\phi^{,\nu}$ gives the Klein-Gordon equation, Eq.~\eqref{KGeqcov}. However, Eq.~\eqref{Tmunumu} are actually four independent equations. Only in the $\gamma = 1$ case they reduce to only one, namely the usual Klein-Gordon equation. Using Eq.~\eqref{KGeqcov} into Eq.~\eqref{Tmunumu} one obtains
\begin{equation}\label{newconstr}
 \phi_{,\nu}\phi_{;\alpha;\beta}\phi^{,\alpha}\phi^{,\beta} - \phi_{,\alpha}\phi^{,\alpha}\phi_{;\nu;\beta}\phi^{,\beta} = 0\;.
\end{equation}
At the background level, being $\phi_0$ dependent only on the time, Eq.~\eqref{newconstr} is identically satisfied. For small perturbations, Eq.~\eqref{newconstr} gives the perturbed Klein-Gordon equation for $\nu = 0$, whereas for $\nu = i$ one has
\begin{equation}\label{newconstr2}
 (1 - \gamma)\ddot{\phi}_0\delta\phi_{,i} = (1 - \gamma)\dot{\phi}_0\delta\dot{\phi}_{,i} - (1 - \gamma)\dot{\phi}_0^2\Phi_{,i}\;.
\end{equation}
For $\gamma \neq 1$ one can cast the above equation as follows:
\begin{equation}\label{newconstr3}
 \left(\frac{\delta\phi_{,i}}{\dot\phi_0}\right)^{\cdot} = \Phi_{,i}\;,
\end{equation}
which appears to be an additional constraint that we have to take into account together with the Einstein's equations.

\section{The scalar field speed of sound in Rastall's theory}

In \cite{Gao:2009me} the authors investigate the case corresponding to $\gamma = 2$, which is able to reproduce the $\Lambda$CDM scenario both at the background and at the perturbative level. A possible explanation for the success of the case $\gamma = 2$ may reside in the fact that in such instance the speed of sound vanishes, as we show now. The speed of sound is defined as the ratio $c_s^2 := \delta p/\delta\rho$, which is gauge-dependent. Therefore, it makes sense to consider its value in the reference frame where the substance whose collapse is being investigated is at rest; we denote such quantity as $\hat{c_s}$.

Following \cite{Bean:2003fb}, we employ the formula
\begin{equation}\label{BDformula}
 \delta p = \hat{c_s}^2\delta\rho + 3aH\rho(1 + w)\left(\hat{c_s}^2 - c_a^2\right)\frac{\theta}{k^2}\;,
\end{equation}
which links the pressure perturbations to the energy density ones, both in a generic gauge, via $\hat{c_s}^2$. In this formula, $c_a^2$ is the adiabatic speed of sound, defined as $c_a^2 := \dot{p}/\dot{\rho}$ and which, for the Rastall scalar field that we are investigating, has the form
\begin{equation}\label{adss}
 c_a^2 = \frac{3H(2 - \gamma)\dot{\phi_0} + 2(3 - 2\gamma)V_{,\phi}}{3H\gamma\dot{\phi_0}}\;,
\end{equation}
where we have employed Eqs.~\eqref{rhoR}, \eqref{pR} and the equation of motion \eqref{KGeqcovbg}. Moreover, in formula \eqref{BDformula}, $w := p/\rho$, $k$ is the wavenumber coming from a normal mode decomposition and $\theta$ is defined via
\begin{equation}\label{theta}
 a(\rho + p)\theta := \partial^i\delta T^0{}_i = \dot{\phi_0}\partial^i\delta\phi_{,i}\;.
\end{equation}
Substituting in Eq.~\eqref{BDformula} the expressions for $\delta\rho$ and $\delta p$, that we have found in Eqs.~\eqref{deltarho} and \eqref{deltap}, and Eqs.~\eqref{adss} and \eqref{theta} we obtain
\begin{eqnarray}
 \gamma\hat{c_s}^2\left[\gamma\dot{\phi_0}\dot{\delta\phi} - \gamma\Phi\dot{\phi_0}^2 + (3 - 2\gamma)V_{,\phi}\delta\phi + 3H\dot{\phi_0}\delta\phi\right] = \nonumber\\
= (2 - \gamma)\left[\gamma\dot{\phi_0}\dot{\delta\phi} - \gamma\Phi\dot{\phi_0}^2 + (3 - 2\gamma)V_{,\phi}\delta\phi + 3H\dot{\phi_0}\delta\phi\right]\;,
\end{eqnarray}
which clearly gives that $\hat{c_s}^2 = (2 - \gamma)/\gamma$ and therefore the case $\gamma = 2$ implies $\hat{c_s}^2 = 0$, which is a favouring case for the collapse.

%%%%%%%%%%%%%%%%%%%%%%%%%%%%%%%%%%%%%%%%%%%%%%%%%%%%%%%%%%%%%%%%%%%%%%%%%%%%%%%%%%%%%%%%%%%%%%%%%%%%%%%%%%%%%%%%%%

\section{Evolution of the gravitational potential}

Following \cite{Mukhanov2005}, we calculate the Einstein tensor from metric \eqref{pertFLRWmet} and combine it with the perturbed stress-energy tensor in Eqs.~\eqref{deltarho}--\eqref{deltap}. In particular, as we implicitly anticipated in writing the perturbed metric \eqref{pertFLRWmet}, since $\delta T^i{}_j \propto \delta^i{}_j$ we have only one gravitational potential. See \cite{Mukhanov2005} for more detail.

We obtain
\begin{eqnarray}
\label{G00} \nabla^2\Phi - 3\mathcal{H}\left(\mathcal{H}\Phi + \Phi'\right) + \gamma\left(\mathcal{H}^2 - \mathcal{H}'\right)\Phi =\nonumber\\ 4\pi G\left[\gamma\phi_0'\delta\phi' + (3 - 2\gamma)a^2V_{,\phi}\delta\phi\right]\;,\\ \nonumber\\
\label{G0i} \mathcal{H}\Phi_{,i} + \Phi'_{,i} = 4\pi G\phi_0'\delta\phi_{,i}\;,\\ \nonumber\\
\Phi'' + 3\mathcal{H}\Phi' + \left(2\mathcal{H}' + \mathcal{H}^2\right)\Phi + (2 - \gamma)\left(\mathcal{H}^2 - \mathcal{H}'\right)\Phi =\nonumber\\ 4\pi G\left[(2 - \gamma)\phi_0'\delta\phi' - (3 - 2\gamma)a^2V_{,\phi}\delta\phi\right]\;,
\end{eqnarray}
where we have also used the background relation $4\pi G\phi_0^{'2} = \mathcal{H}^2 - \mathcal{H}'$.

Now we reduce the above system in two different ways which, however, will give different results unless we choose $\gamma = 1$. Let us employ a normal mode decomposition. The second equation can thus be written as $\mathcal{H}\Phi + \Phi' = 4\pi G\phi_0'\delta\phi$ and $\nabla^2 = -k^2$.

Now, if we sum or subtract the first equation with the third, use $\mathcal{H}\Phi + \Phi' = 4\pi G\phi_0'\delta\phi$ in order to eliminate $\delta\phi$ and the equation of motion \eqref{KGeqcovbgconf} in order to eliminate $\phi_0''$ we obtain:
\begin{eqnarray}
\label{Eq1} \Phi'' + 3\mathcal{H}\Phi' + \left(2\mathcal{H}' + \mathcal{H}^2\right)\Phi =\nonumber\\ \frac{2 - \gamma}{\gamma}\left[-k^2\Phi - 3\mathcal{H}\left(\mathcal{H}\Phi + \Phi'\right)\right] - \frac{2V_{,\phi}a^2}{\gamma\phi_0'}(3 - 2\gamma)\left(\mathcal{H}\Phi + \Phi'\right)\;,\\ \nonumber\\
\label{Eq2} \Phi'' + 3\mathcal{H}\Phi' + \left(2\mathcal{H}' + \mathcal{H}^2\right)\Phi =\nonumber\\ -k^2\Phi - 3\mathcal{H}\frac{2 - \gamma}{\gamma}\left(\mathcal{H}\Phi + \Phi'\right) - \frac{2V_{,\phi}a^2}{\gamma\phi_0'}(3 - 2\gamma)\left(\mathcal{H}\Phi + \Phi'\right)\;.
\end{eqnarray}
These equations can be identical only if $\gamma = 1$, which is the General Relativity limit of Rastall's theory.

Actually, these equations could be consistent also if $k = 0$, but what does this mean? Going back to the Einstein equations, before the normal modes decomposition, $k = 0$ would mean $\nabla^2\Phi = 0$, which implies that $\Phi$ should be a homogeneous field, i.e. $\Phi = \Phi(\eta)$. Note that any spatial linear dependence for $\Phi$ would be unacceptable since the field would diverge at infinity.

But if $\Phi$ is homogeneous, then we are not facing real perturbations. In fact, it is sufficient in metric \eqref{pertFLRWmet} to redefine the time and the scale factor as follows:
\begin{equation}
 d\bar{t}^2 = [1 + 2\Phi(t)]dt^2\;, \qquad \bar{a}^2(t) = a^2(t)[1 - 2\Phi(t)]\;,
\end{equation}
and we obtain again the FLRW metric in the usual reference frame, with $\bar{t}$ the cosmic time.

Therefore we conclude that a consistent perturbation theory, formulated in the newtonian gauge, is possible only for $\gamma = 1$.

Is such result compatible with the new constraint \eqref{newconstr3} that we have showed to exist in Rastall's theory, for $\gamma \neq 1$? The answer is yes, since combining Eqs.~\eqref{G00} and \eqref{G0i} it is possible to obtain Eq.~\eqref{newconstr3}. Therefore, the latter is not an actual constraint: it is already embedded in Einstein's equations.

\section{The role of a fluid component}

The result found in the previous section appears to be different if we introduce another component together with the Rastall scalar field. Indeed, let us consider a perfect fluid with equation of state $p = w\rho$ and $w$ constant. We write its total (i.e. background plus perturbed) stress-energy tensor as follows:
\begin{eqnarray}
\label{rhof} T^0{}_0 &=& \rho(1 + \delta)\;,\\
 T^0{}_i &=& -\rho(1 + w)v_i\;,\\
\label{pf} T^i{}_j &=& -(p + \delta p)\delta^i{}_j\;,
\end{eqnarray}
where $\delta := \delta\rho/\rho$ is the usual density contrast, $\delta\rho$ is the density perturbation, $\delta p$ is the pressure one and $v_i$ is the velocity. We assume adiabatic perturbations, i.e. $\delta p = c_{\rm s}^2\delta\rho$, where $c_{\rm s}^2 = w$.

Employing again the normal mode decomposition, we rewrite the system of linearised Einstein equations as follows:
\begin{eqnarray}
\label{G00new} -k^2\Phi - 3\mathcal{H}\left(\mathcal{H}\Phi + \Phi'\right) + \gamma\left(\mathcal{H}^2 - \mathcal{H}'\right)\Phi = 4\pi G a^2\delta\rho_\phi + 4\pi G a^2\delta\rho\;,\\ \nonumber\\
\label{G0inew} k\left(\mathcal{H}\Phi + \Phi'\right) = 4\pi G k \phi_0'\delta\phi + 4\pi G\rho(1 + w)v\;,\\ \nonumber\\
\label{Gijnew} \Phi'' + 3\mathcal{H}\Phi' + 3\mathcal{H}^2\Phi -\gamma\left(\mathcal{H}^2 - \mathcal{H}'\right)\Phi = 4\pi G a^2\delta p_\phi + 4\pi G a^2 \delta p\;,
\end{eqnarray}
where we have defined:
\begin{equation}
 \delta\rho_\phi := \frac{1}{a^2}\gamma\phi_0'\delta\phi' + (3 - 2\gamma)V_{,\phi}\delta\phi\;,
\end{equation}
\begin{equation}
 \delta p_\phi := \frac{1}{a^2}(2 - \gamma)\phi_0'\delta\phi' - (3 - 2\gamma)V_{,\phi}\delta\phi\;,
\end{equation}
and $v$ is the velocity potential defined by $v_i = -v_{,i}/k$.

With the fluid variables and the relation $\delta p = c_{\rm s}^2\delta\rho$ we have a total of four unknowns $(\Phi, \delta\phi, \delta\rho, v)$ but only 3 equations. Therefore, it is impossible to obtain again a constraint as strong as $\gamma = 1$. 

Let us investigate in some detail the coupled system fluid plus Rastall scalar field. Multiplying Eq.~\eqref{G00new} by $2 - \gamma$, Eq.~\eqref{Gijnew} by $\gamma$ and subtracting the two we obtain:
\begin{eqnarray}
 \gamma\Phi'' + 6\mathcal{H}\Phi' + 6\mathcal{H}^2\Phi - 2\gamma\left(\mathcal{H}^2 - \mathcal{H}'\right)\Phi + (2 - \gamma)k^2\Phi = \nonumber\\ -8\pi G a^2(3 - 2\gamma)V_{,\phi}\delta\phi + 4\pi G a^2\left(\gamma c_{\rm s}^2 + \gamma -2\right)\delta\rho \;.
\end{eqnarray}
From this equation we eliminate $\delta\phi$ with the help of Eq.~\eqref{G0inew}, obtaining
\begin{eqnarray}
\label{Phieqnew} \gamma\Phi'' + 6\mathcal{H}\Phi' + 6\mathcal{H}^2\Phi - 2\gamma\left(\mathcal{H}^2 - \mathcal{H}'\right)\Phi + (2 - \gamma)k^2\Phi = \nonumber\\ -2a^2(3 - 2\gamma)V'\left[\frac{\mathcal{H}\Phi + \Phi'}{\phi_0^{'2}} - \frac{4\pi G a^2\rho(1+w)v}{k\phi_0^{'2}}\right]\nonumber\\ 
+ 4\pi G a^2\rho\left(\gamma c_{\rm s}^2 + \gamma -2\right)\delta\;,
\end{eqnarray}
where we have used $V_{,\phi} = V'/\phi'$. We now assume the fluid to satisfy its own energy-momentum conservation, separately from the scalar field, in order to gain one more equation necessary to solve the system. From $T^\mu{}_{\nu;\mu} = 0$, for the fluid component only, we get:
\begin{eqnarray}
 \delta' &=& -(1 + w)(kv - 3\Phi')\;,\\
v' &=& -\mathcal{H}(1 - 3w)v + \frac{kc_{\rm s}^2}{1 + w}\delta + k\Phi\;.
\end{eqnarray}
Now we have to specify the background evolution, i.e. the function $\mathcal{H}$. Its general form is
\begin{equation}
 \frac{\mathcal{H}^2}{a^2} = \frac{8\pi G}{3}\left[\rho + \frac{\gamma\phi_0^{'2}}{a^2} + (3 - 2\gamma)V\right]\;.
\end{equation}
In order to simplify considerably Eq.~\eqref{Phieqnew}, we assume that the potential is a constant, i.e. $V' = 0$. Therefore, it is going to play the role of an effective cosmological constant. If the potential is a constant, it turns out from the Klein-Gordon equation \eqref{KGeqcovbg} that
\begin{equation}
 \phi_0' = u_0 a^{-(3 - \gamma)/\gamma}\;,
\end{equation}
and the Friedmann equation takes on the following form:
\begin{equation}\label{FriedEqVconst}
 \frac{\mathcal{H}^2}{H_0^2a^2} = \Omega_0 a^{-3(1 + w)} + \Omega_V + \Omega_{u_0}a^{-6/\gamma}\;,
\end{equation}
with the definitions
\begin{equation}
 \Omega_V := \frac{8\pi G(3 - 2\gamma)V}{3H_0^2}\;, \qquad \Omega_{u_0} := \frac{8\pi G\gamma u_0^2}{3H_0^2}\;.
\end{equation}
Finally, trading the conformal time for the scale factor, we write the coupled system of Einstein equations plus the fluid equations as
\begin{eqnarray}
\label{Phieqnew2} \gamma\Phi_{aa} + \left[\gamma\frac{\dot{\mathcal{H}}}{\mathcal{H}} + \frac{6 + \gamma}{a}\right]\Phi_{a} + \left[\frac{2\gamma\dot{\mathcal{H}}}{\mathcal{H}a} + \frac{6 - 2\gamma}{a^2}\right]\Phi + (2 - \gamma)\frac{k^2}{\mathcal{H}^2a^2}\Phi = \nonumber\\ \frac{3H_0^2}{2\mathcal{H}^2}\Omega_0 a^{-3(1 + w)}\left(\gamma c_{\rm s}^2 + \gamma -2\right)\delta\;,\\
\delta_a = -(1 + w)\left(\frac{kv}{\mathcal{H}a} - 3\Phi_a\right)\;,\\
v_a = -\frac{1}{a}(1 - 3w)v + \frac{kc_{\rm s}^2}{(1 + w)\mathcal{H}a}\delta + \frac{k}{\mathcal{H}a}\Phi\;,
\end{eqnarray}
where the subscript $a$ denotes derivation with respect to the scale factor. 

We start the evolution from $a_i = 10^{-3}$ and choose as initial conditions $\Phi'_i = v_i = 0$, $\Phi_i = -\delta_i = -1$.

With the choice $w = 0$ and $\gamma = 2$, Eqs.~\eqref{FriedEqVconst} and \eqref{Phieqnew2} reproduce the same dynamics of the $\Lambda$CDM model. It is curious that this seems to happen only when a standard fluid is added to the Rastall scalar field.
\begin{figure}[htbp]
 \includegraphics[width=0.45\columnwidth]{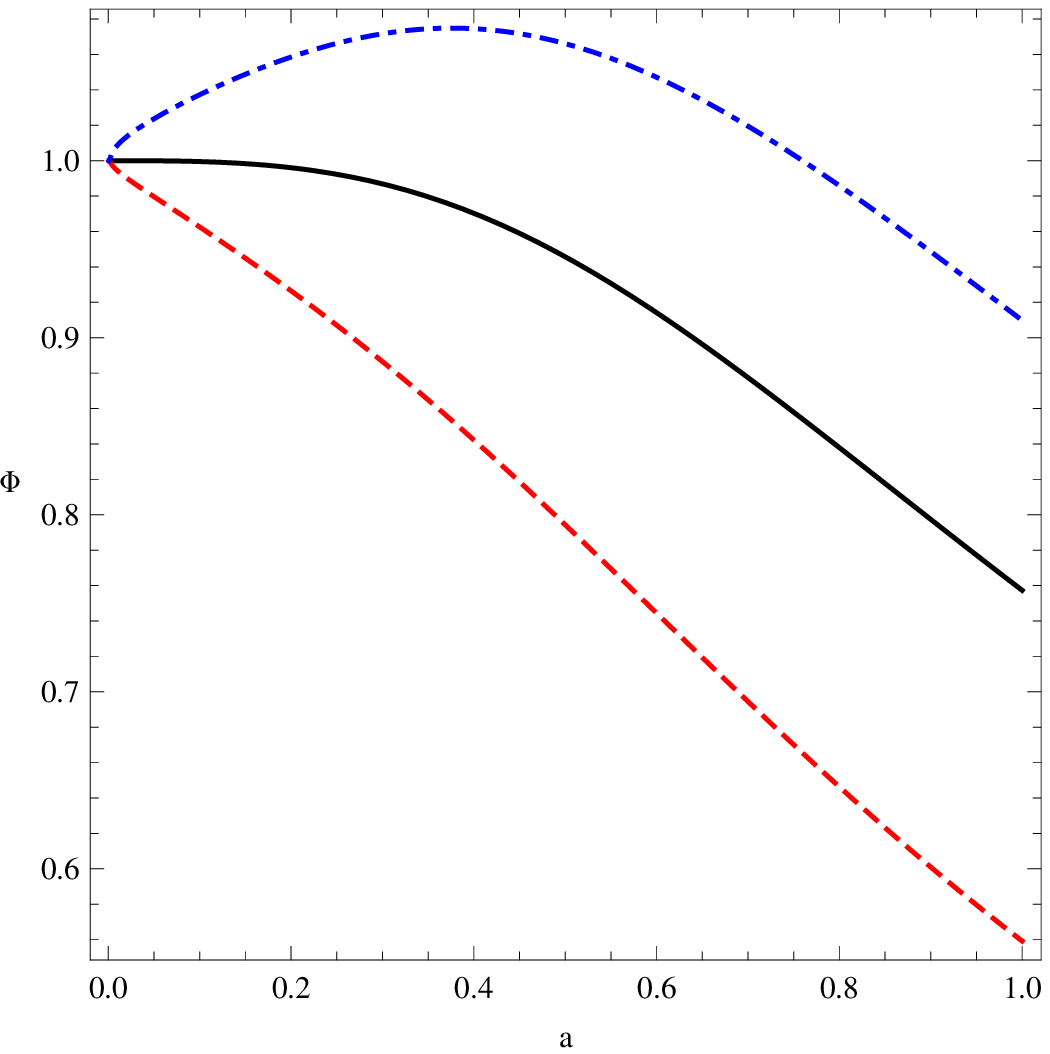}\hspace{0.05\columnwidth}\includegraphics[width=0.45\columnwidth]{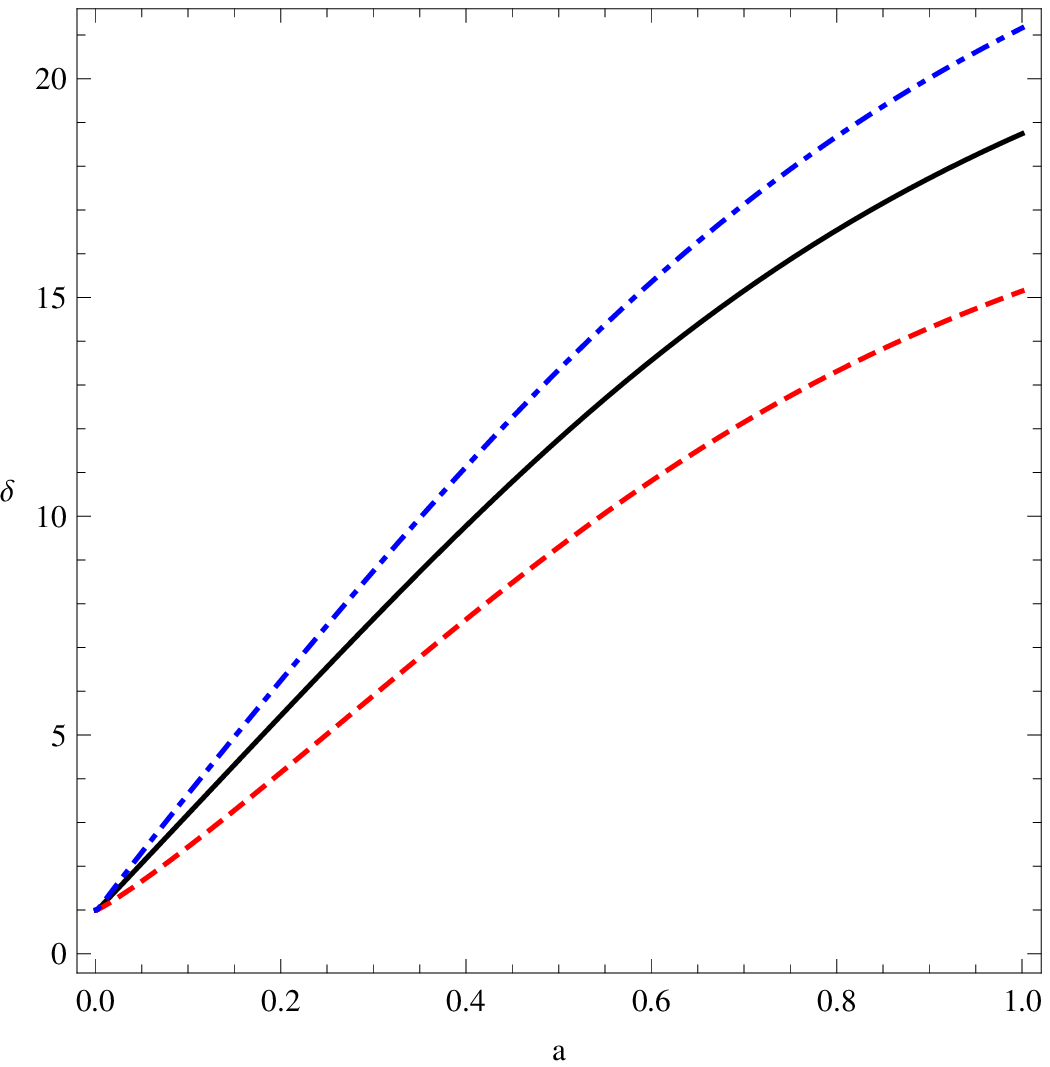}\\
\caption{Evolution of $\Phi$ and $\delta$ for the choice $w = 0$ and $\Omega_0 = 0.04$. The black solid lines represent the case $\gamma = 2$, i.e. the $\Lambda$CDM one. The red dashed lines correspond to $\gamma = 1.9$ whereas the blue dot-dashed ones to $\gamma = 2.05$. We have chosen a scale $k = 10^{-3}$ h Mpc$^{-1}$.}
\label{fig1}
\end{figure}
In \figurename{ \ref{fig1}} and \ref{fig2} we plot the evolution of $\Phi$ and $\delta$ for the choice $w = 0$ and $\Omega_0 = 0.04$. That is, we are assuming that the perfect fluid under consideration is a baryon component. We have also chosen a scale $k = 10^{-3}$ h Mpc$^{-1}$.
\begin{figure}[htbp]
 \includegraphics[width=0.45\columnwidth]{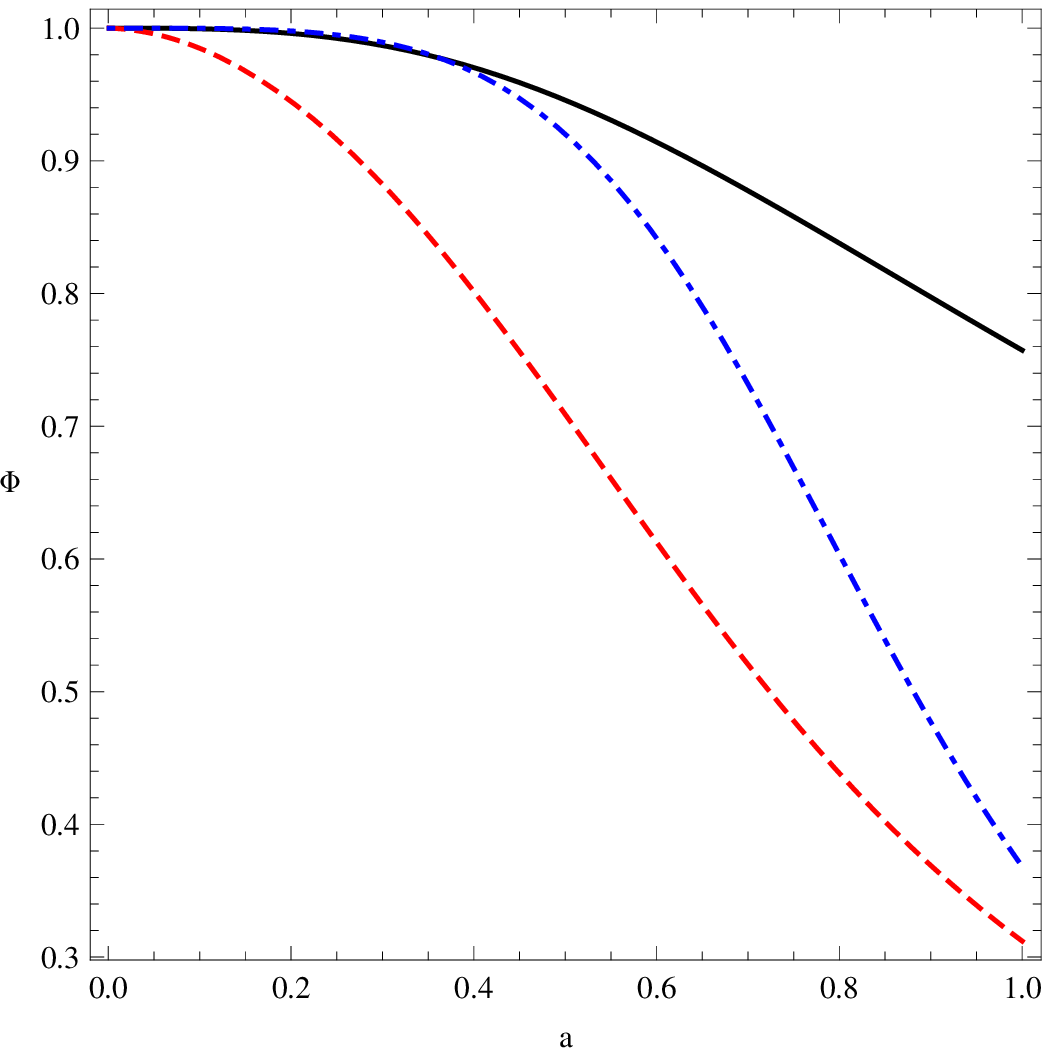}\hspace{0.05\columnwidth}\includegraphics[width=0.45\columnwidth]{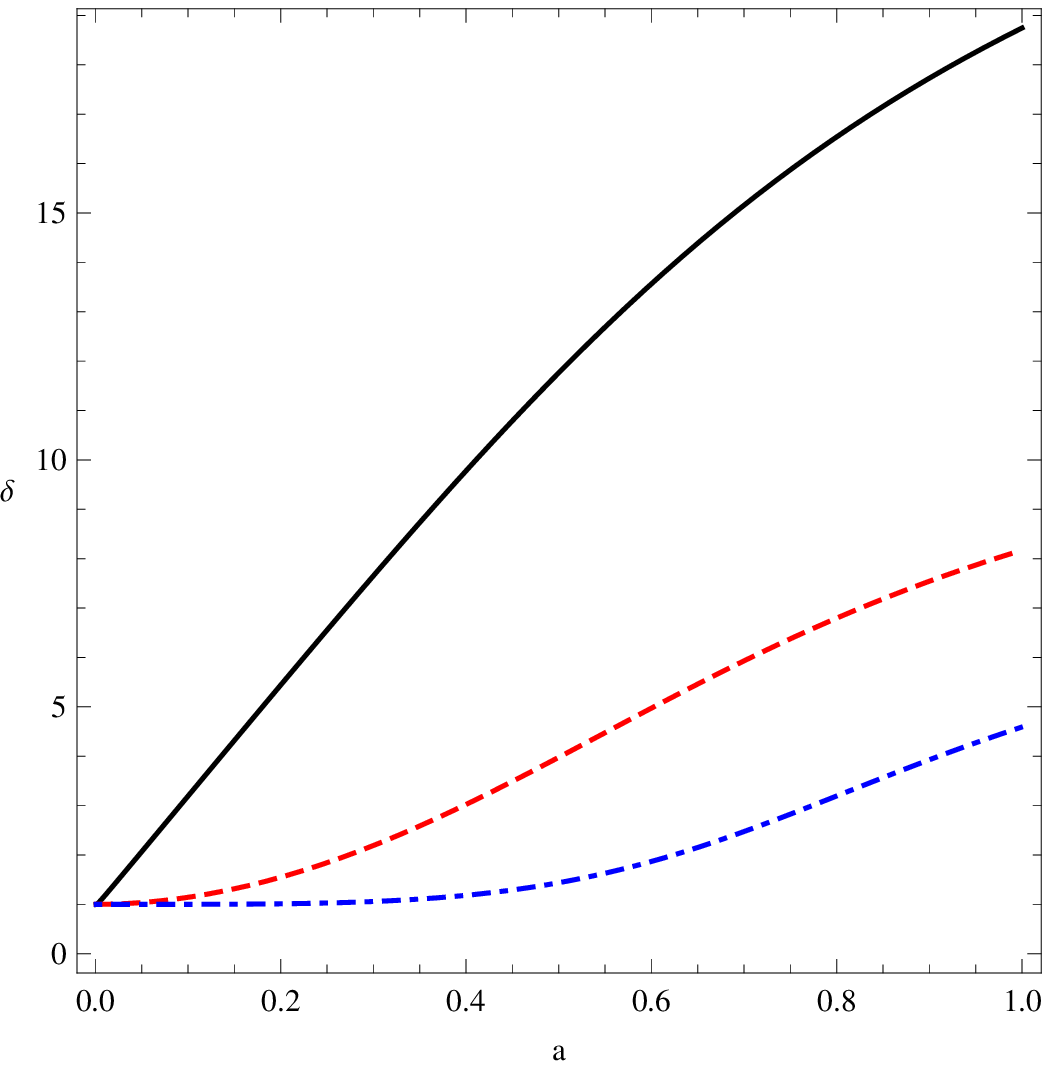}\\
\caption{Evolution of $\Phi$ and $\delta$ for the choice $w = 0$ and $\Omega_0 = 0.04$. The black solid lines represent the case $\gamma = 2$, i.e. the $\Lambda$CDM one. The red dashed lines correspond to $\gamma = 1.5$ whereas the blue dot-dashed ones to $\gamma = 1$. We have chosen a scale $k = 10^{-3}$ h Mpc$^{-1}$.}
\label{fig2}
\end{figure}
It seems that, for $\gamma > 2$, the growth of the density contrast of the fluid component is enhanced. This is probably due to the fact that $\hat{c}_{\rm s}^2$ becomes negative and therefore the collapse of the scalar field is unimpeded. For $\gamma < 2$ the growth of $\delta$ is sensibly hampered. It is curious in \figurename{ \ref{fig2}} how the gravitational potential suffers a larger decrease for the case $\gamma = 1.5$ than for the $\gamma = 1$ one. We would have expect the contrary, since for $\gamma = 1$ the speed of sound is $\hat{c}_{\rm s}^2 = 1$, whereas for $\gamma = 3/2$ the speed of sound is $\hat{c}_{\rm s}^2 = 1/3$, i.e. smaller. Note that such discrepancy is not present in the plots for $\delta$, i.e. the growth for $\gamma = 1.5$ is larger than the one for $\gamma = 1$. Therefore, such effect is probably due to the different background evolution.

\begin{figure}[htbp]
\begin{center}
 \includegraphics[width=0.55\columnwidth]{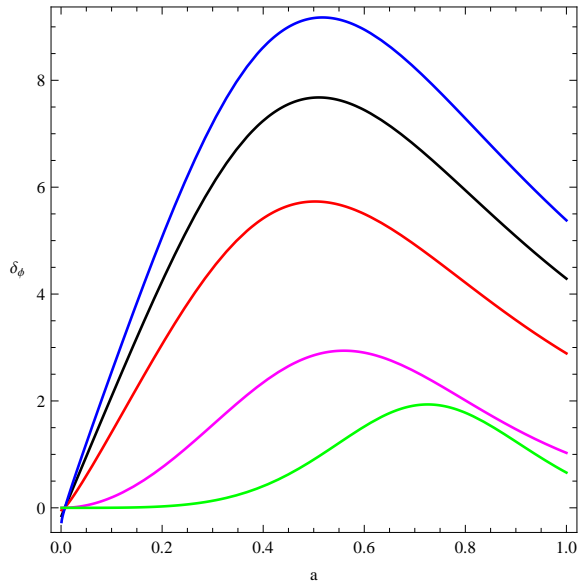}\\
\end{center}
\caption{Evolution of $\delta_\phi$ for the choice $w = 0$ and $\Omega_0 = 0.04$. From top to bottom: $\gamma = 2.05, 2, 1.9, 1.5, 1$. We have chosen a scale $k = 10^{-3}$ h Mpc$^{-1}$.}
\label{fig3}
\end{figure}

For completeness, we display here also the evolution of $\delta_\phi := \delta\rho_\phi/\rho_\phi$, which can be easily calculated from Eq.~\eqref{G00new}, once we know $\Phi$ and $\delta$. The results are plotted in \figurename{ \ref{fig3}}.

Some comments about Eq.~\eqref{newconstr3} are in order, since the latter establishes a strong connection between perturbations in the scalar field and in the gravitational potential. For a single scalar field component, Rastall's theory reduces to General Relativity, thus everything runs as in the standard lore. On the other hand, the coupling with matter brings new features. Here also comes the complexity given by the new conservation law of Rastall's theory, which admits many consistent alternatives. If the matter component conserves separately (as we have investigated in this paper), Eq.~\eqref{newconstr3} remains untouched. However, there are examples in which it may change, e.g. the case in which matter exchanges energy with the scalar field or the one investigated in \cite{Batista:2011nu}, where two fluid components were considered, one of them still conserving separately, whereas the conservation of other depending on the curvature. In this case, it is not difficult to show that relation~\eqref{newconstr3} gives place to
\begin{equation}
\left(\frac{\delta\phi_{,i}}{\dot\phi_0}\right)^{\cdot} = \Phi_{,i} - \frac{1}{2}\left(\delta\rho_{,i} - 3\delta p_{,i}\right)\;,
\end{equation}
rendering the situation somewhat different. In \cite{Batista:2011nu} other possible couplings between the two components are also evoked, which may lead to other variants for Eq.~\eqref{newconstr3}.

Even if our interest in this work is the late-times universe, we may ask for potential consequences of Eq.~\eqref{newconstr3} and its possible variants for the primordial spectrum. One point is that Rastall's scalar model requires another component in order to make sense. But, an adiabatic primordial spectrum can be naturally implemented mainly if the matter component is subdominant with respect to the scalar component. The isocurvature component can also be implemented, in principle, since it requires a zero total (scalar plus matter) density fluctuation $\delta\rho_{\rm tot} = 0$, and this even if the relation~\eqref{newconstr3} still holds, as in the case where the other component conserves separately, without direct interaction with the scalar field. When other types of interactions between both components are considered, as in the equation above, the isocurvature perturbation can still exist. But, in general, the detailed predictions for the spectrum must differ from the standard cases, mainly in the isocurvature case. This may open a new path of investigation concerning the specific predictions of Rastall's theory for the primordial spectrum of perturbations.

Equation~\eqref{newconstr3} also reminds the relation between the spatial curvature and the inflaton field, in the standard inflationary scenario \cite{Liddle:2000cg}.  However, such connection still has to be investigated in detail, in order to understand how inflation could be implemented into Rastall's theory.

\section{Conclusions}

The scalar formulation of Rastall's theory of gravity may allow a consistent unification of dark matter and dark energy for the background evolution of the universe. We have shown in this paper that, on the other hand, its single component version (with the non-canonical Rastall scalar field as the only matter content) is perturbatively inconsistent: the compatibility of the perturbed equations requires a homogeneous gravitational potential, which is equivalent to a redefinition of the background functions and not to real perturbations. The Rastall scalar theory may admit consistent perturbative scenario if another (canonical) fluid component is added.
\par
For a two-fluid model, we numerically evaluate the behaviour of the gravitational potential and that of the density contrast for the scalar field component. For some cases, as $\gamma \sim 1.5$, a behaviour very similar to that obtained in the General Relativity case with a quintessence field is obtained. Although this does not represent an exhaustive study of the Rastall two-component model, the results here reported indicate that consistent scenarios may emerge from a Rastall unification model for dark energy. We hope to present in future a more exhaustive study with a detailed comparison with cosmological observational data.

%%%%%%%%%%%%%%%%%%%%%%%%%%%%%%%%%%%%%%%%%%%%%%%%%%%%%%%%%%%%%%%%%%%%%%%%%%%%%%%%%%%%%%%%%%%%%%%%%%%%%%%%%%%%%%%%

\section*{Acknowledgements} We thank CNPq (Brazil) for partial financial support. We also acknowledge the anonymous referee for his useful remarks and suggestions. We are indebted to W.~Zimdahl for many enlightening discussions. 

%%%%%%%%%%%%%%%%%%%%%%%%%%%%%%%%%%%%%%%%%%%%%%%%%%%%%%%%%%%%%%%%%%%%%%%%%%%%%%%%%%%%%%%%%%%%%%%%%%%%%%%%%%%%%%%

\bibliographystyle{model1-num-names}

\end{document}